\documentclass[preprint,double-spaced,showpacs,amsmath,amssymb]{revtex4}\textwidth=14.5cm \textheight=24cm
\usepackage{amssymb,graphicx}
\usepackage{dcolumn}

\def\P3{{\cal P}_t}
\def\J3{{\cal J}}
\def\T3{{\cal T}}

\def\beq{\begin{equation}}
\def\eeq{\end{equation}}
\def\bar{\begin{array}[b]}
\def\barc{\begin{array}}
\def\bart{\begin{array}[t]}
\def\ear{\end{array}}

 1  1
 1 
scaled\magstephalf 
 
scaled\magstephalf

\begin{document}
\thispagestyle{empty} \vspace*{0.5 cm}
\begin{center}
{\bf Neutrino flavor dynamics in post-bounce supernovae :
the role of the electron component.}
\\
\vspace*{1cm} {M. Baldo and V. Palmisano}\\
\vspace*{.3cm}
{\it INFN, Sezione di Catania}\\
{\it Via S. Sofia 64, I-95123, Catania, Italy} \\
\vspace*{.6cm} \vspace*{1 cm}
\end{center}

\begin{center}{\bf ABSTRACT} \\\end{center}

In this paper we study the dynamics of flavor transformation for neutrinos propagating in the very dense environment of astrophysical compact objects
as Type II supernova in post collapse phase and proto-neutron stars. The analysis is based on the generalized Boltzmann equation, which incorporates
the neutrino-neutrino interaction. We focus the analysis on the possible collective flavor dynamics, which can displays bipolar and synchronized
flavor oscillations, and the associated transition from single-angle to multi-angle regime. To solve the kinetic equations we use an expansion in
Legendre polynomials and a two-flavor scheme. On the basis of the numerical simulations we argue that neutrino flavor dynamics is suppressed by the
presence of the electron component up to a certain distance from the neutrino sphere, where a transition to multi-angle regime occurs. This distance
will move towards the neutrino sphere as the neutrino emission epoch proceeds. \vskip 0.3 cm

PACS :
14.60.Pq ,  
26.50.+X ,  
26.60.-c ,  
95.30.Cq ,  
97.60.Jd .  

\section{Introduction}

Today it is commonly accepted that the study of neutrinos from core collapse supernovae is an important contribution to the development of our
knowledge about elementary particle physics and also to the understanding of the evolution mechanism of astrophysical sources
\cite{beth,taka,thom,bur1,full,sigl}. In fact the supernovae in the post bounce stage are substantially similar to a black body emitting neutrinos of
all three flavors, whose propagation in the very dense stellar material is highly sensitive to parameters such as mass hierarchy or the mixing
angles. At large distance from the core $r \gg 10^3$ km the celebrated MSW resonance \cite{bil1,bil2,bald,bal1,fog1,fog2} is able to convert the
flavour of each neutrino. This phenomenon was decisive to solve the neutrino puzzle in the solar case \cite{wolf,smir}, but it is also relevant for
the neutrino signals that can come from a supernova \cite{bald,bal1}. \par At much shorter distances from the core of the star, $r \sim$ $10 - 150$
km, the concentration of produced neutrinos is so high that self-interactions is not any more negligible. Numerical simulations in recent studies
show that in this situation interesting effects emerge, all characterized by the collective behavior of neutrinos and antineutrinos which are coupled
regardless of their energy or propagation direction \cite{bal2,fog3,kos2,raf1,raf2,raf3,raf4,qian}. These effects are different from ordinary
oscillations in matter because they involve colliding neutrinos with a cross section that depends on the angle between their momenta. These
particular features introduce a set of coupled non-linear equations and makes prohibitive an exact solution of the kinetic equations in the general
case \cite{raf5,raf6}. However reliable approximations have been developed, which have lead to a detailed studies of the neutrino flavor dynamics.

\par

One of the main results \cite{raf1,raf2,raf3,raf4,qian} of these studies is the prediction that at a certain distance from the neutrino sphere a
transition from coherent flavor oscillations to incoherent ones should occur. The coherent oscillations correspond to flavour oscillations that are
in phase among the different direction along which neutrino are propagating, while in the incoherent ones the different directions are decoupled and
the flavour oscillations are out of phase. Another predicted process is a sharp transition from the synchronized oscillations to the so called
``bipolar'' ones, that should mainly correspond to the mutual conversion of neutrino-antineutrino pairs of one flavour to another one. This second
transition is somehow independent from the first one, since it can occur even when a single-angle description is adopted \cite{raf3}. These processes
are strongly affected by the matter physical conditions, besides the neutrino and antineutrino flavor compositions. In particular the presence of
electrons can reduce the effective mixing angle in the matter to a very small value. For high enough electron density it is expected that all flavor
oscillation/conversion processes will be hindered by the increase of the electron effective potential on the neutrinos due to neutral weak current
interaction.

\par

In this paper we analyse the behavior of neutrino flavor content by Boltzmann-like equations, following the formulation of Strack and Burrows
\cite{stra}. Outside the neutrino sphere neutrinos flow essentially freely, where both neutrino and electron densities are decreasing. We focus the
analysis on the effects of the electron component and we consider the physical conditions at different distances from the neutrino sphere as the
after-bounce stage evolves. The flavor evolution is simulated under realistic assumptions about the matter physical parameters within the
uncertainties of the problem.

\section{The formalism and numerical simulations}

\par

To simulate the neutrino flavor evolution we use the generalized Boltzman-like kinetic equation that incorporate both neutrino-neutrino interaction
and the coupling with the star background matter. The latter is characterized by a density profile that evolves in time along the neutrino emission
epoch. This evolution has been described in supernovae simulations \cite{pons1,pons2,janka} and it can be estimated by static calculations of
proto-neutron stars at different entropy content. As an illustration we report in Fig.~\ref{fig:fig1} a sample of these matter density profiles at
different entropy of the envelope calculated following the scheme of Ref.~\cite{burg}. The calculation assume neutrino trapping, thermodynamic and
chemical local equilibrium, and an entropy per particle $S = 1$ in the core and $S = 1, 2, 3, 4$ in the envelope. The lowest entropy profiles should
correspond to the latest stage of the neutrino emission period. The results are in line with dynamical simulations of the Livermore group, see Fig. 1
of Ref.~\cite{schi}. Despite the quasi-static calculations do not extend to the more far tails of the profiles and the two sets of calculations use
two different types of scheme, they show a very similar trend  and their overall evolution suggests the range of density values that the profile
should cover during the neutrino emission epoch. We will assume an electron fraction of 0.4, a value close to the one found in the static
calculations \cite{burg}.

\par

The kinetic equation is the space-time evolution equation for the Wigner-Ville distribution in phase space, which is the basic quantity of the
scheme. It is a quasi-probability distribution introduced to study quantum corrections to classical statistical mechanics which links the wave
function, that appears in Schr\"odinger's equation, to a probability distribution in phase space. In second quantization formalism it can be defined
as

\begin{equation}\label{wigner-density}
\rho \left( {{\bf{r}},{\bf{p}},t} \right) = \int {\frac{{d^3
{\bf{r'}}}}{{\left( {2\pi \hbar } \right)^3 }}e^{ -i{\bf{pr'}}} \psi
^\dag  \left( {{\bf{r}} - \frac{1}{2}{\bf{r'}},t} \right)} \psi
\left( {{\bf{r}} + \frac{1}{2}{\bf{r'}},t} \right),
\end{equation}

\noindent where $\psi ^\dag$ and $\psi$ are creation and annihilation operators.

As discussed in Ref. \cite{stra} it is possible to develop a formalism to describe transport processes in terms of Wigner functions, which is able to
incorporate the quantum phenomenon of flavor conversion of neutrinos. The basic equations of the theory can be derived heuristically merging
Boltzmann equation with Heisenberg equation. Taking the matrix elements in Fock space, considering for simplicity only two neutrino flavors, in terms
of the matrix elements of the density matrix~(\ref{wigner-density}) the complete evolution equation is ~\cite{raf3,este,stra}

\begin{equation}
\begin{array}{*{20}{l}}
{{\bf{\dot f}} = \Omega \left( {E,t} \right)\cdot{\bf{f}}}\\
{}\\
\begin{array}{l}
\Omega \left( {E,t} \right) = {\Omega _{vacuum}}\left( E \right) + {\Omega _{matter}}\left( t \right) + {\Omega _{\nu \nu }}\left( t \right)

\end{array}
\end{array}
\end{equation}

In this equation  ${\bf f}$ is a 2$\times$2 matrix in flavor space. The diagonal matrix elements are real valued quantities that have the meaning of
phase space densities, while off-diagonal terms are complex quantities with the  meaning of macroscopic overlap functions. The quantity
$\Omega_{vacuum}$ is the vacuum mixing Hamiltonian. Besides this, the equation includes in general the term corresponding to the interaction with
matter $\Omega_{matter}$ and the term ${\Omega _{\nu \nu }}$ that describes the interaction of neutrinos and anti-neutrinos among themselves.
 It is convenient to introduce the real and imaginary part of off diagonal macroscopic overlap $f_r  $ and $f_i $ and write the
generalized Boltzmann equations in terms of real quantities only. For details and the explicit expressions see ref. \cite{stra}.

We start our analysis by considering a simplified scheme where the density both of neutrinos and of matter are uniform and constant in time. We study
the evolution of the flavour content in a two-flavour framework in the "solar" scheme, with the mixing angle parameter $\sin(2\theta) \,=\, 0.92 $
and mass splitting $\Delta m^2 \,=\, 7.59\times 10^{-5}$eV$^2$. For simplicity the neutrino are taken mono-energetic, at energy $E_\nu \,=\, 10$ MeV.
Under these physical conditions the relevance of the electron component can be directly and quantitatively determined. The kinetic equation can be
written in this case by neglecting the space derivatives. The term proportional to the momentum derivative can be in general neglected because it is
only due to the gravitational field. Since we are interested in the neutrino outflow above the neutrino sphere, where neutrino are essentially free
flowing, we do not include neither the neutrino-matter collisions nor the absorption/emission processes. To simulate the neutrino emission from the
neutrino sphere we assume that the initial angular distribution is non-zero only in the forward direction and, for simplicity, uniform. We take
"symmetric" initial conditions, for which the  "electron" neutrinos and anti-neutrinos are equal in number, while  "muon" neutrinos or anti-neutrinos
are absent. Then, at the fixed total neutrino density $\rho_\nu \,=\, 10^{32}$/cm$^3$, we tune the electron density in a range of values compatible
with the profiles of Fig.~\ref{fig:fig1}. The results are reported in Fig.~\ref{fig:fig2}, where the left column refers to direct hierarchy and the
right column to the inverse one. In the ordinate is reported the relative abundance of electron neutrinos with respect to the total neutrino density.
At low enough electron density a sharp transition is apparent after a very short time. It has to be stressed that in this short time $t$ neutrino can
propagate by a distance shorter than the characteristic distance over which the neutrino and matter density variation is relevant. In the particular
case of  Fig.~\ref{fig:fig2}c this distance is $d \,=\, ct \,\approx\, 3 $Km. As the density increase there is a critical value $\rho_c$ above which
the transition disappears. We have checked that at density higher than this critical value the signal is flat for times of the order of several
seconds. After a finer search, we found that the critical density can be estimated as $\rho_c \approx 2\times 10^7$g/cm$^3$. To elucidate the nature
of this transition we have calculated the angular distributions of the electron neutrino content as a function of time, see Fig.~\ref{fig:fig3}.
Above the critical density the angular distribution does not change with time, while for lower density there is a sharp transition from a uniform
angular distribution to an irregular one, exactly at the time where the electron neutrino component drops. We can then identify this transition as a
transition from a single-angle evolution to a multi-angles one. In the inverse hierarchy the situation is similar, except that the transition occurs
at slightly shorter time. In these calculations we have assumed a focusing of the angular distributions according to the prescription of e.g. ref.
\cite{duan}, see Fig. 1 of this work. This is apparent in Fig.~\ref{fig:fig3}. The focusing has been estimated by assuming a neutrino sphere of $10$
Km and a distance of $15$ Km. The results do not depend critically on this particular choice. Indeed, a similar behavior is found at the larger
distance of $50$ Km.

\par

Let us consider what can be happening of the flavor contents as the neutrinos propagate from the neutrino sphere outwards. At the border of the
neutrino sphere the electron density is expected to be much higher than the critical value discussed above. Then the flavor distribution cannot vary
in time. At higher distance the electron density can drop below the critical value, then the transition from the single-angle to the multi-angles
evolution should occur within a short time and within a distance short enough to assume the neutrino density as constant. This justify the assumption
of constant density in the calculations. However, as we will see, this is not a crucial requirement. In order to check further this picture we have
repeated the calculations assuming that the matter density varies in time to simulate the evolution of the matter density profile as the neutrino
propagate outwards. We take an exponential shape $\exp(-t/\tau_p)$ with different time scale $\tau_p$. As reported in  Fig. \ref{fig:fig4}, the
transition indeed occurs as the matter density crosses the critical value discussed above, while the neutrino flavor density matrix remains constant
before this point. These profiles start at a density of $10^{11} g/$cm$^3$, which is a typical value at the border of the neutrino sphere, and drop
in a time scale $\tau_p$. The distance $L \,=\, c\tau_p$ can be taken as the length scale which characterizes the matter density drop above the
neutrino sphere. The chosen values of $\tau_p$ correspond to $L$ values typical of the matter profile within $1-30$ s after bounce, as it can be seen
in Fig. \ref{fig:fig1} as compared with the Fig. 1 of Ref.~\cite{schi}. It can be seen from Fig. \ref{fig:fig4} that the transition point is shifted
upwards as the density profile becomes less steep and it follows the shift of the point where the critical density is reached.  Despite the
simplified picture, the results give clear evidence that the transition from single-angle to multi-angle regime cannot occur before the critical
density is reached, and indeed it should occur just after this point. As a further check we report in Fig. \ref{fig:fig5} the real and imaginary
parts of the off-diagonal elements of the density matrix $\mathbf{f}$, which describes the flavor mixing, in the case with $t_{cr} \,=\, 2\times
10^{-5}$. As one can see they remain zero until the density is close to the critical one. This completes the evidence that the flavor dynamics is
fully blocked before the critical density is reached.

\par

Of course the neutrino density profile is changing with the radial distance $R$, approximately as $1/R^2$. This reinforces the blocking of the
neutrino flavor dynamics, since then the effect of electrons will be stronger, and the point of the transition will be further shifted upwards.

In agreement with ref. \cite{este} we found also that if the neutrino flow is asymmetric, the picture changes drastically and as the asymmetry
reaches a critical value, that is estimated to be not larger than $0.1$, the flavor neutrino dynamics is suppressed. These findings, together with a
more extensive report of the results will be presented elsewhere.

\section{Discussion and conclusion}

For a given flavor composition and symmetric neutrino-antineutrino flavor content, we have estimated the critical matter density below which the
transition from single-angle to multi-angle behavior can be expected and the time scale of the transition for a wide set of density values. The
method is based on the expansion of the Boltzmann equations in Legendre polynomials. The transition is apparent in the angular distribution of the
flavor contents, which is independent of time before the transition and becomes irregular at the transition.  The focusing of the neutrino flow at
different distances modifies both the critical density and the time scale. Once a matter density profile is given, one could get an estimate of the
region where the transition should first occur. Since the density profile changes with neutrino emission time, this region will move with time. In
general the matter density profile is uniformly decreasing along the emission time, therefore the region for the transition will move toward the
neutrino sphere. Furthermore since the neutrino flavor modulations are more apparent near the neutrino sphere, the neutrino signal should be stronger
at the later stage of the neutrino emission epoch. This type of analysis is usually performed by introducing the so called ``polarization vector''
\cite{raf3,este}, whose three-dimensional evolution in time (or distance) gives an overall view of the flavor dynamical evolution. We preferred to
use directly the full angular distribution of the flavor content since it gives a more detailed description of the flavor structure.

\par

\section*{Acknowledgements}

\begin{figure}[t]
\begin{center}
\includegraphics[width=0.9\textwidth]{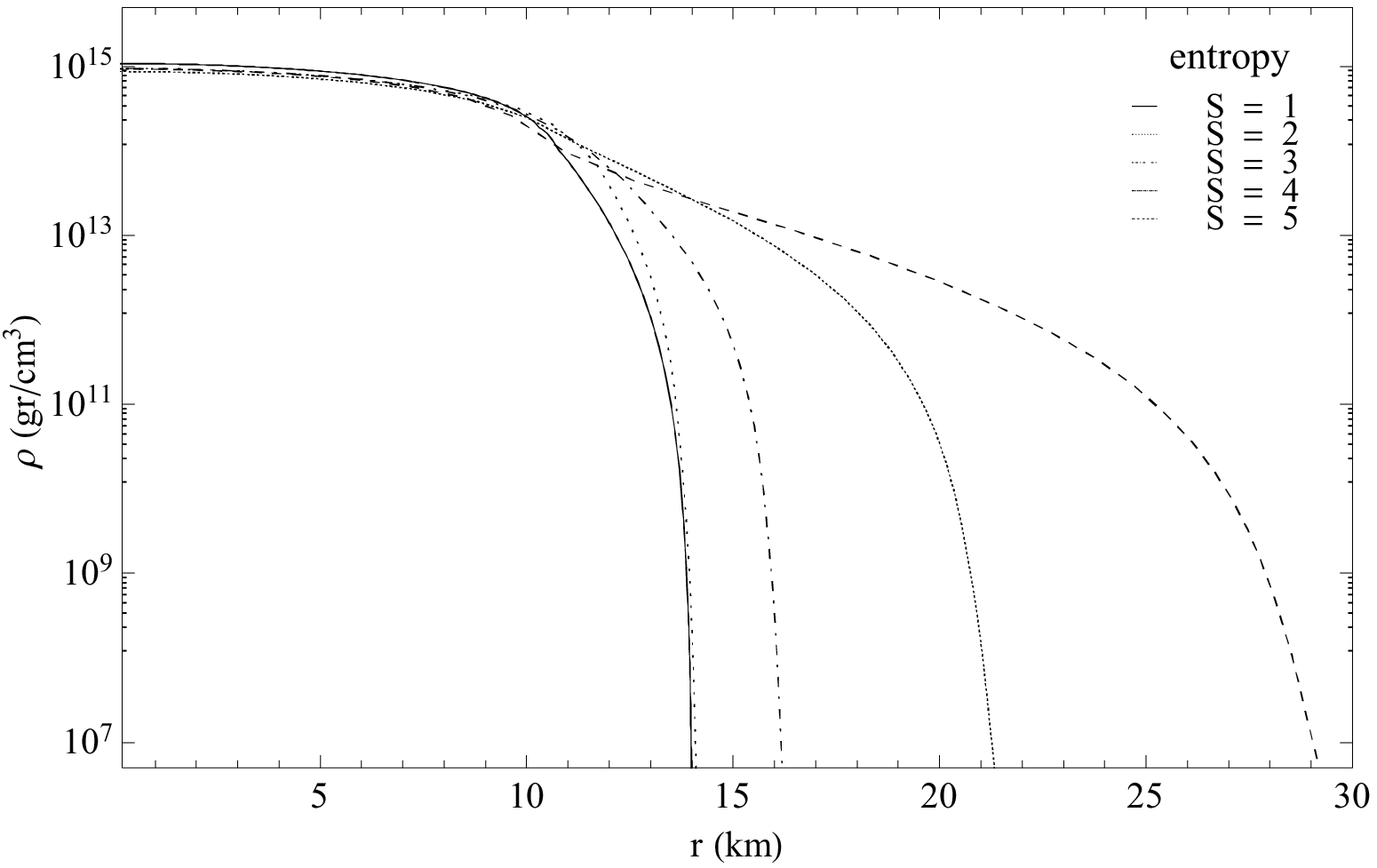}\caption{Baryon density profiles of a proton-neutron star of mass
$M/M_\odot = 1.4$ for different entropy. The core entropy is fixed at S = 1 while the envelope entropy  is taken, from the lower to the higher
profiles, as S = 1, 2, 3, 4, 5.} \label{fig:fig1}
\end{center}
\end{figure}

\begin{figure}[t]
\begin{center}
\includegraphics[width=1.15\textwidth]{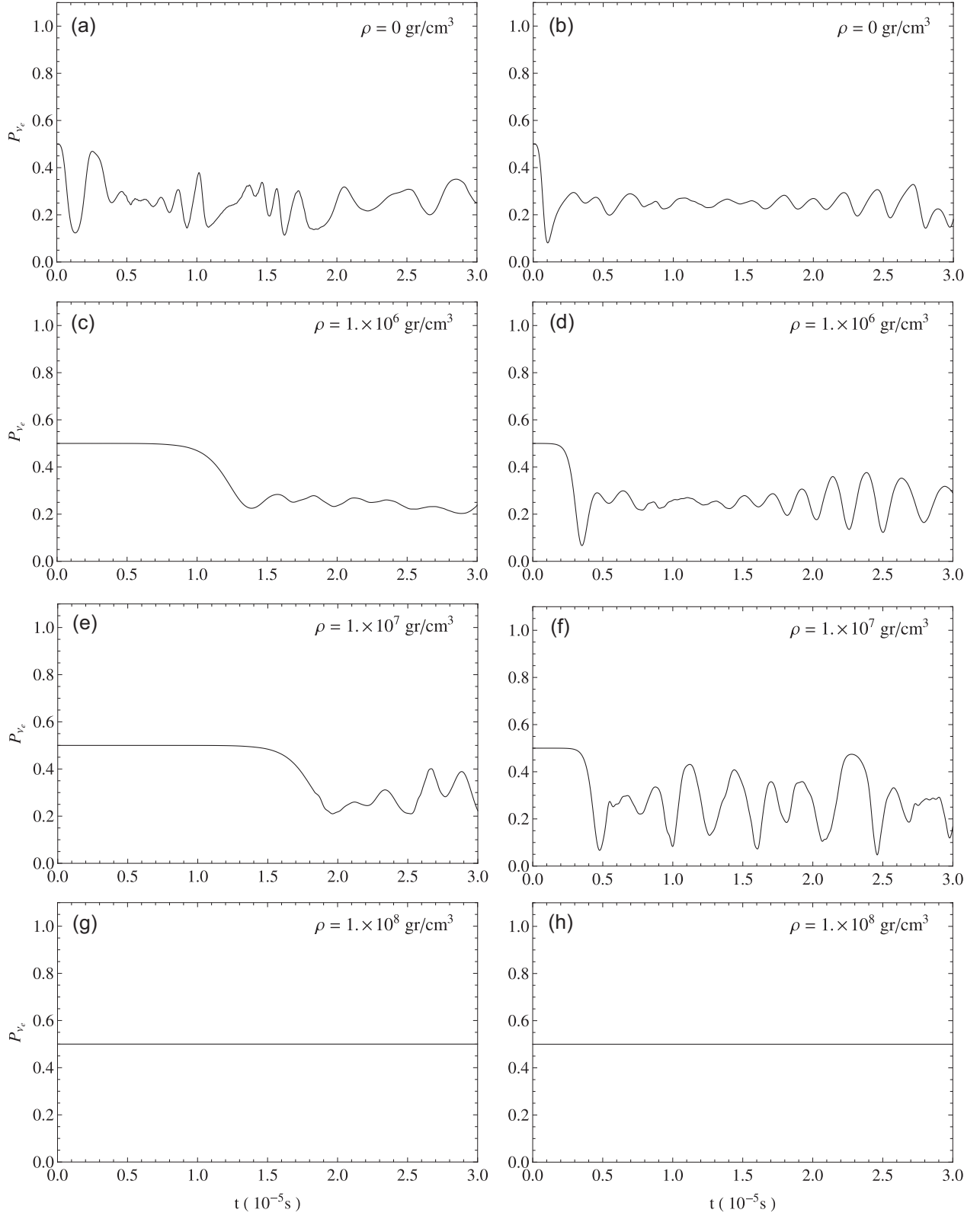}
\caption{Time evolution of $\nu_e$ fraction at radial distance of $15$ km and different values of the matter density $\rho$. Energy is fixed to $10$ MeV and
neutrino density to $10^{32}$ cm$^{-3}$. Time is in unit of 10$^{-5}$s. Left: normal hierarchy. Right: inverted hierarchy. }
\label{fig:fig2}
\end{center}
\end{figure}

\begin{figure}[t]
\begin{center}
\includegraphics[width=0.8\textwidth]{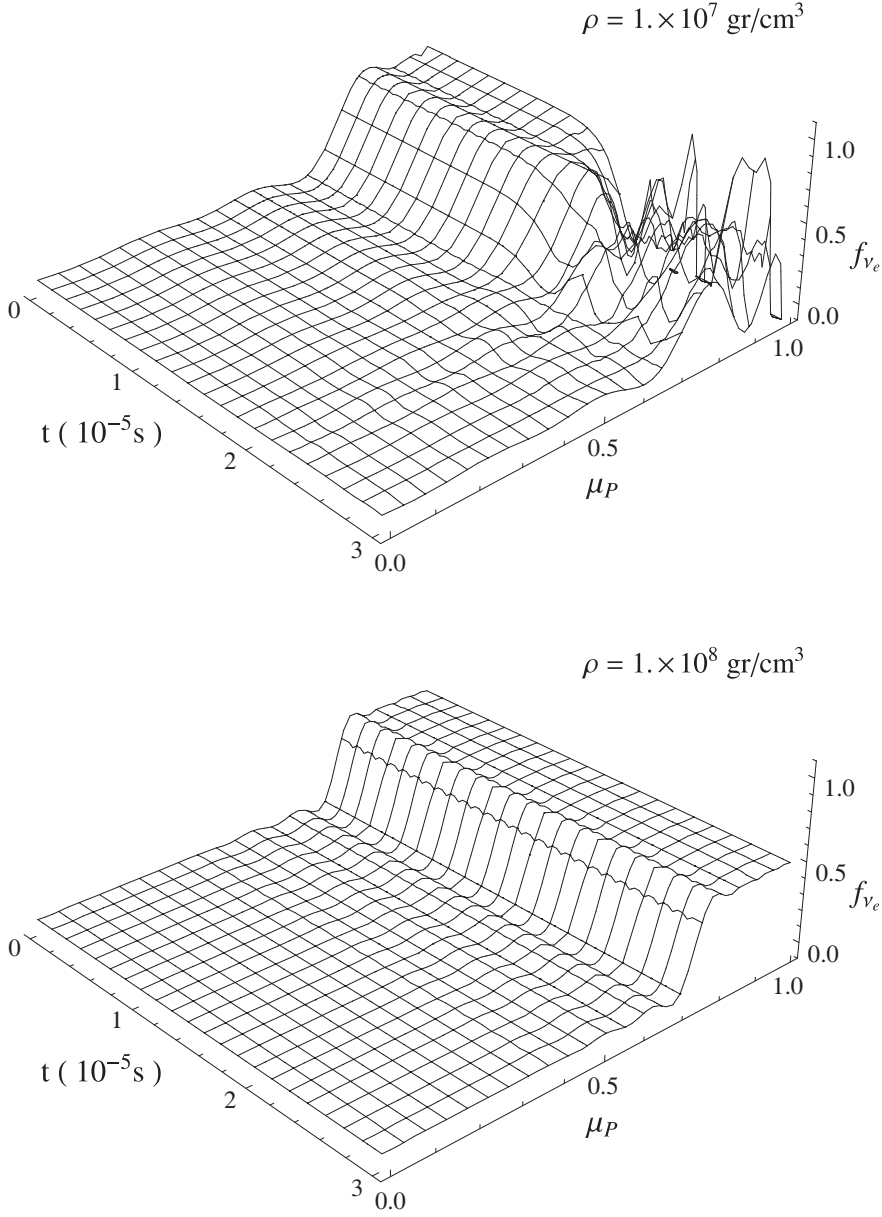}
\caption{Angular distribution of $\nu_e$ content in function of time with same parameter of Fig.~\ref{fig:fig2} in normal hierarchy. Quantity
$f_{\nu_e}$ is normalized to total neutrino density and $\mu_p=\cos(\theta_p)$, with $\theta_p$ the angle with respect to the radial direction. Upper
plot: density matter fixed to $10^7$ gr/cm$^3$. Lower plot: density matter fixed to $10^8$ gr/cm$^3$. } \label{fig:fig3}
\end{center}
\end{figure}

\begin{figure}[t]
\begin{center}
\includegraphics[width=0.75\textwidth]{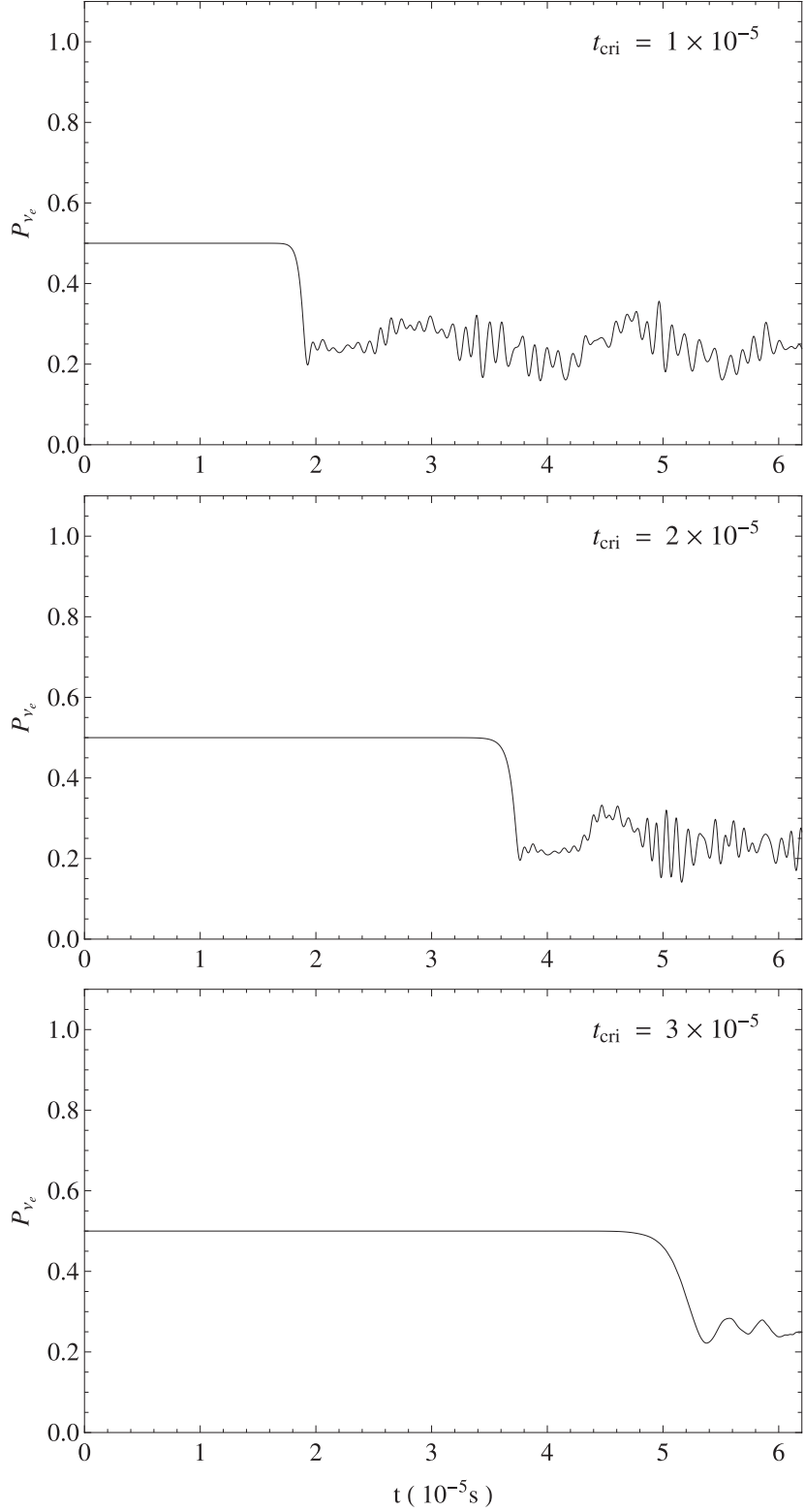}
\caption{The electron neutrino fraction as a function of time with a time dependent matter density profile and for neutrino density fixed at $10^{32}$ cm$^{-3}$. The labels indicate the time at which the matter density passes through the critical value $\rho_c \approx 2\times 10^7$g/cm$^3$.}   \label{fig:fig4}
\end{center}
\end{figure}

\begin{figure}[t]
\begin{center}
\includegraphics[width=0.75\textwidth]{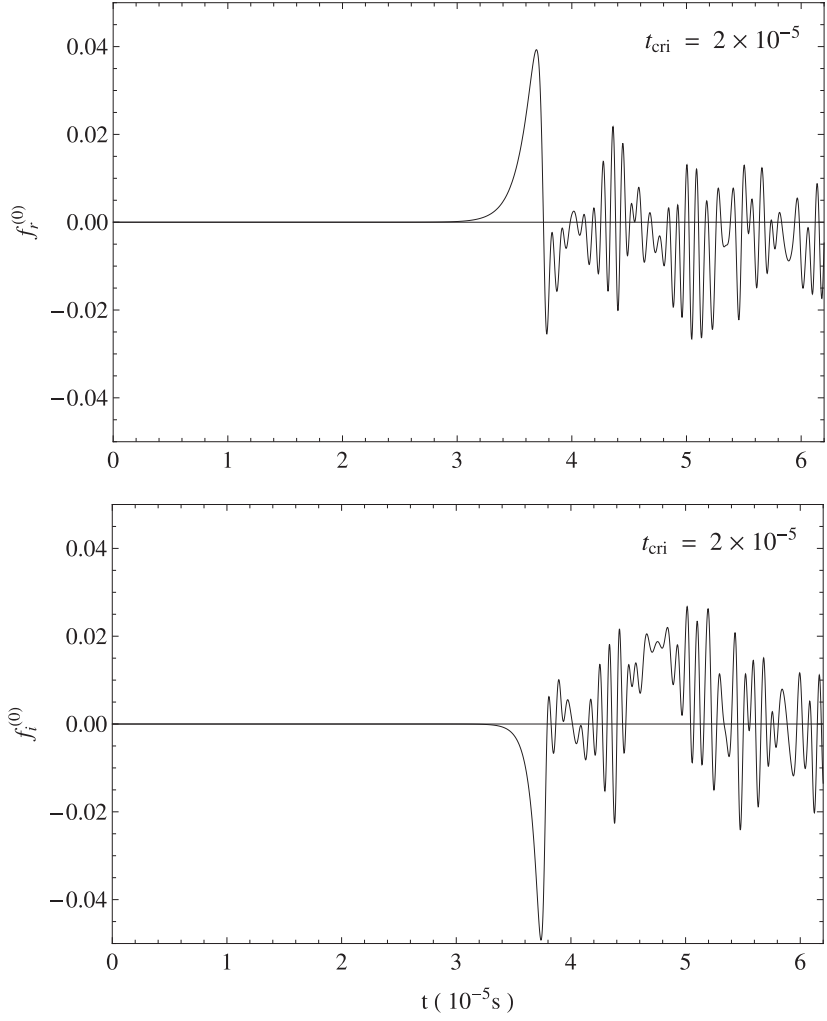}
\caption{The real and imaginary parts $f_r$ and $f_i$ of the off-diagonal matrix element in the density matrix as function of time. The labels
indicate the time at which the matter density passes through the critical value $\rho_c \approx 2\times 10^7$g/cm$^3$.}   \label{fig:fig5}
\end{center}
\end{figure}

\addcontentsline{toc}{section}{} We thank Dr. G. F. Burgio  for providing us the density profiles reported in Fig.~\ref{fig:fig1}.

\vfill\eject

\vfill\eject

\vfill\eject

\vfill\eject

\vfill\eject

\vfill\eject

\end{document}